\documentclass[twocolumn,aps,prb,groupedaddress,showpacs]{revtex4}
\usepackage[cp1252]{inputenc}
\usepackage{textcomp}
\usepackage{amsmath}
\usepackage{amssymb}
\usepackage{graphicx}
\usepackage{esint}

\makeatletter

\@ifundefined{textcolor}{}
{%
 \definecolor{BLACK}{gray}{0}
 \definecolor{WHITE}{gray}{1}
 \definecolor{RED}{rgb}{1,0,0}
 \definecolor{GREEN}{rgb}{0,1,0}
 \definecolor{BLUE}{rgb}{0,0,1}
 \definecolor{CYAN}{cmyk}{1,0,0,0}
 \definecolor{MAGENTA}{cmyk}{0,1,0,0}
 \definecolor{YELLOW}{cmyk}{0,0,1,0}
}

\usepackage{xcolor}

\newcommand{\be}{\begin{equation}}
\newcommand{\ee}{\end{equation}}
\newcommand{\bea}{\begin{eqnarray}}
\newcommand{\eea}{\end{eqnarray}}

\newcommand{\la}{\langle}
\newcommand{\ra}{\rangle}

\def\H-AltDM{{\cal H}_{\text{\tiny Alt-DM}}}

\makeatother

\begin{document}

\title{Long-range spin chirality dimer order in the Heisenberg chain with
modulated Dzyaloshinskii-Moriya interactions}
\author{N.\ Avalishvili G.I.\ Japaridze}
\affiliation{\small{\sl Ilia State University, Faculty of Natural
Sciences and Medicine, 0162, Tbilisi, Georgia}} \affiliation{\small
{\sl Andronikashvili Institute of Physics, 0177, Tbilisi, Georgia}}
\author{G.L.\ Rossini}
\affiliation{\small {\sl IFLP-CONICET and Departamento de
F\'{i}sica, Universidad Nacional de La Plata,\\
 CC 67 1900 La Plata, Argentina}}

\begin{abstract}

The ground state phase diagram of a spin $S=1/2$  $XXZ$ Heisenberg
chain with spatially modulated Dzyaloshinskii-Moriya (DM)
interaction
$$
{\cal H}= \sum_{n}J\left[\left(S^{x}_{n}S^{x}_{n+1}
+S^{y}_{n}S^{y}_{n+1}+\Delta
S^{z}_{n}S^{z}_{n+1}\right)+(D_{0}+(-1)^{n}D_{1})\left(S^{x}_{n}S^{y}_{n+1}
-S^{y}_{n}S^{x}_{n+1} \right) \right]
$$
is studied using the continuum-limit bosonization approach and
extensive density matrix renormalization group computations. It is
shown that the effective continuum-limit bosonized theory of the
model is given by the double frequency sine-Gordon model (DSG) where
the frequences i.e.\ the scaling dimensions of the two competing
cosine perturbation terms depend on the effective anisotropy
parameter $\gamma^{\ast}=J\Delta /\sqrt{J^2+D_{0}^{2}+D_{1}^{2}}$.
Exploring the ground state properties of the DSG model we have shown
that the zero-temperature phase diagram contains the following four
phases: (i) the ferromagnetic phase at $\gamma^{\ast}\leq -1$; (ii)
the gapless Luttinger-liquid (LL) phase at $-1<\gamma^{\ast}<
\gamma^{\ast}_{c1}=-1/\sqrt{2}$; (iii) the gapped composite (C1)
phase characterized by coexistence of the long-range-ordered (LRO)
dimerization pattern $\epsilon \sim (-1)^{n} ({\bf S}_{n}{\bf
S}_{n+1})$ with the LRO alternating spin chirality pattern $\kappa
\sim (-1)^{n}\left(S^{x}_{n}S^{y}_{n+1} -S^{y}_{n}S^{x}_{n+1}
\right)$ at $\gamma^{\ast}_{c1}<\gamma^{\ast} <\gamma^{\ast}_{c2}$;
and (iv) at $\gamma^{\ast}
>\gamma^{\ast}_{c2}>1$ the gapped composite (C2) phase characterized
in addition to the coexisting spin dimerization and alternating
chirality patterns, by the presence of LRO antiferromagnetic order.
The transition from the LL to the C1 phase at $\gamma^{\ast}_{c1}$
belongs to the Berezinskii-Kosterlitz-Thouless universality class,
while the transition at $\gamma^{\ast}=\gamma^{\ast}_{c2}$ from C1
to C2 phase is of the Ising type.

\end{abstract}

\date{Today}

\maketitle

\maketitle

\section{\bf Introduction}

Quantum spin chains continue to be the subject of intensive studies
because they serve as interesting model systems to explore strongly
correlated quantum order in low dimensional magnetic systems
[\onlinecite{Mikeska_Kolezhuk,Broholm_et_al_08,Vasiliev_Volkova_18}].
A significant fraction of current research is focused on studies of
helical structures and chiral order in the frustrated quantum
magnetic systems
[\onlinecite{Zviagin,Derzhko_1,Derzhko_2,Oshikawa_Affleck,YuLu_2003,Mahdavifar_08,Aristov_Maleev_00,Tsvelick_01,Starykh_08,Garate_Affeck_10,Starykh_17,Fazio_14,Mila_et-al_11,Langari_09,Mahdavifar_14a,Ising_DM_1,Ising_DM_2}].
The key couplings, responsible for stabilization of non-collinear
magnetic configurations in these systems, is the
Dzyaloshinskii-Moriya (DM) interaction [\onlinecite{DMI-1,DMI-2}]
\begin{equation}\label{DM-term}
{\cal H}_{DM}=\sum_{n}{\bf D}(n)\cdot[{\bf S}_{n}\times{\bf
S}_{n+1}]\, ,
\end{equation}
where ${\bf D}(n)$ is an axial DM vector. The DM interaction
corresponds to the antisymmetric part of exchange interaction
between spin located on neighboring sites $n$ and $n+1$, it appears
in a systems with broken inversion symmetry due to the spin-orbit
coupling and was first introduced by I.\ Dzyaloshinskii on the
grounds of general symmetry arguments [\onlinecite{DMI-1}]. Later,
the spin-orbit coupling as the microscopic mechanism of the
antisymmetric exchange interaction has been identified by T.\ Moriya
[\onlinecite{DMI-2}].

Although the study of helical structures in antiferromagnets counts
more then half of century [\onlinecite{Dzyaloshinskii_64}], the
research activity in the field of one and quasi-one-dimensional
spin-1/2 chains with DM interaction remain persistent and high
during the last three decades.
Effects caused by the uniform DM term or by the pure staggered DM
interaction on the ground state properties of the $S=1/2$ Heisenberg
chain were considered within the framework of Bethe-Ansatz solvable
models [\onlinecite{Zviagin}], as well as using the exactly solvable
limiting cases such as the $XY$ chain with uniform DM couplings
[\onlinecite{Derzhko_1}]. Magnetic properties of the isotropic and
anisotropic ($XXZ$) Heisenberg chain with staggered
[\onlinecite{Oshikawa_Affleck,YuLu_2003,Mahdavifar_08}] and uniform
[\onlinecite{Tsvelick_01,Starykh_08,Garate_Affeck_10,Starykh_17}] DM
interaction has been considered  using the continuum-limit
bosonization approach and numerical treatment. Recently more exotic
extended versions of the one-dimensional Heisenberg model, such as
the completely anisotropic spin-1/2 XYZ model with DM interaction
[\onlinecite{Fazio_14}] and the Delta-chain model with DM
interaction [\onlinecite{Mila_et-al_11}] have been studied using the
density-matrix renormalization group algorithm (DMRG) and a
finite-size scaling analysis. In last years, using the exact
diagonalization technique, a special attention has been given to the
studies of the ground state phase diagram of finite spin chains with
DM interaction based on calculation of the entanglement, for the
Heisenberg [\onlinecite{Langari_09,Mahdavifar_14a}], Ising
[\onlinecite{Ising_DM_1}] and bond-alternating Ising model
[\onlinecite{Ising_DM_2}].

Generally, in a chain, vectors ${\bf D}(n)$ may spatially vary both
in direction and magnitude, however,  the symmetry  restrictions
based on the properties of real solid state materials usually rule
out most  of  the possibilities and  confine  the majority of
theoretical discussion to two principal cases -- uniform DM
interaction, ${\bf D}$ vector remains unchanged over the system
[\onlinecite{Zviagin,Aristov_Maleev_00,Tsvelick_01,Starykh_08}] and
the case of staggered DM interaction,  with antiparallel orientation
of ${\bf D}$ on adjacent bonds
[\onlinecite{Oshikawa_Affleck,YuLu_2003}]. Exception is only the
Ref. [\onlinecite{Derzhko_2}] where the $XY$ spin chain with random
changes in the sign of DM interactions was studied.
\begin{figure}[hb]
\begin{center}
\includegraphics[width=8.0cm,angle=0]{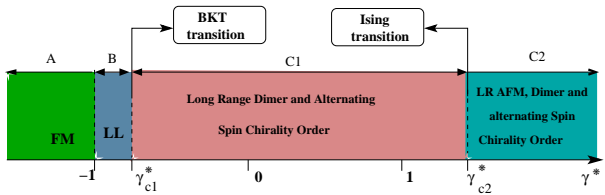}
\caption{ Sketch of the ground state phase diagram of the spin S=1/2
Heisenberg chain with modulated DM interaction.} \label{fig:Fig1}
\end{center}
\end{figure}

However, recently it has been demonstrated that DM interaction can
be efficiently tailored with an substantial efficiency factor by
structural modulations [\onlinecite{Geometric_tailor_DM}] or by
external electric field
[\onlinecite{EF_Enhanc_DM_1,EF_Enhanc_DM_2,EF_Enhanc_DM_3}]. This
unveils the possibility not only to control DM interaction and
magnetic anisotropy via the electric field or other controllable
ways, but also opens a possibility to consider effect of more
general spatially modulated DM interaction on properties of the spin
chain. External electric field induced modulation of the DM
interaction can be realized in spin-driven chiral multiferroic (MF)
systems [\onlinecite{MF_Materials}], effectively coupling the
ferroelectric polarization with the applied external electric field
[\onlinecite{Katsura_Nagaosa_05}]. These studies became very actual
in last years, in particular in the context of materials useful for
electric field controlled quantum information processing
[\onlinecite{EFC_QI_Processing}].

In the present work we study the effect of the alternating
Dzyaloshinskii-Moriya (DM) interaction on the ground state phase
diagram of the spin-1/2 Heisenberg chain. Because the DM term breaks
the global spin rotation symmetry, we consider the Hamiltonian
$$
{\cal H}= {\cal H}_{XXZ} + {\cal H}_{DM}
$$
where
\begin{eqnarray}\label{Hamilt_XXZ}
&&\hspace{-5mm}{\cal H_{XXZ}} =
\sum_{n}\left[J(S^{x}_{n}S^{x}_{n+1}+S^{y}_{n}S^{y}_{n+1})+J_z\,
S^{z}_{n}S^{z}_{n+1}\right]
\end{eqnarray}
is the Hamiltonian of an anisotropic Heisenberg chain and ${\cal
H}_{DM}$ is the DM term given in Eq.\ (\ref{DM-term}). In what
follows we choose the ${\bf D}(n)$ vector orientated, in the spin
space, along the ${\hat z}$ axis ${\bf D}(n)=(0,0,D(n))$ and take
\begin{eqnarray}
D(n)&=&D_{0}+(-1)^{n}D_{1}\, .
\end{eqnarray}

Our main objective is to show that the spatial modulation of the DM
interaction leads to dramatic change of the ground state phase
diagram of the system, opens a gap in a wide area of the parameter
range and also changes the nature of quantum phase transition in the
long-range-ordered antiferromagnetic phase. Results are summarized
in the Fig.\ \ref{fig:Fig1}. The ground state phase diagram is
divided into four sectors depending on the value of the effective
exchange anisotropy parameter
$\gamma^{\ast}=J_{z}/\sqrt{J^{2}+D_{0}^{2}+D_{1}^{2}}$. The
$\gamma^{\ast}=-1$ point corresponds to the transition into the
ferromagnetically ordered phase (sector A). The gapless
Luttinger-liquid phase is shrunk up to a narrow region between $-1
<\gamma^{\ast}<\gamma^{\ast}_{c1}=-\sqrt{2}/2$ (sector B). At
$\gamma^{\ast}=\gamma^{\ast}_{c1}$ the
Berezinskii-Kosterlitz-Thouless (BKT) phase transition takes the
system into the composite (C1) gapped phase  characterized by the
coexistence of long-range ordered (LRO) alternating spin
dimerization pattern
$$
\epsilon(n) = \la{\bf S}_{n}\cdot {\bf S}_{n+1}\ra \sim const
+(-1)^{n} \epsilon
$$
coexisting with long-range alternating pattern of the spin chirality
vector
$$
\kappa^{z}_{n} = \la\left[\,{\bf S}_{n}\times {\bf
S}_{n+1}\,\right]_{z}\ra \sim  const +(-1)^{n} \kappa\, .
$$
Finally, at $\gamma^{\ast}=\gamma^{\ast}_{c2}>1$ there is an Ising
type phase transition into the other  composite (C2) gapped phase,
characterized by the coexistence of long-range dimerization,
chirality and antiferromagnetic
$$
\la S_{n}^{z} \ra \sim const +(-1)^{n}m
$$
modulations.

The outline of the paper is as follows. In Sec.\ II we consider the
exactly solvable limit case of the $XY$ chain with alternating DM
interaction. In Sec.\ III using the gauge transformation we gauge
out the DM coupling and obtain an effective $XXZ$ spin-chain
Hamiltonian with alternating transverse exchange. In Sec.\ IV we
construct the weak-coupling bosonized version of the effective
Hamiltonian and discuss ground state phase diagram. In Sec.\ V we
present extensive numerical results supporting the bosonization
predictions.  Finally, a brief summary is presented in Sec. VI.

\section{\bf The XX chain with alternating DM interaction.}
\label{sec:TheModel}

In this Section we consider the exactly solvable case of a XX chain
with alternating DM interaction. It is instructive to start from the
full Hamiltonian
\begin{eqnarray}\label{Hamiltonian_XXZ_Alt_DM}
{\cal H} &=&\sum_{n}\Big[\frac{J}{2}
\left(S^{+}_{n}S^{-}_{n+1}+S^{-}_{n}S^{+}_{n+1}\right)+J_{z}\,S^{z}_{n}S^{z}_{n+1} \nonumber\\
&+&\frac{i}{2}(D_{0}+(-1)^{n}D_{1})\left(S^{+}_{n}S^{-}_{n+1}
-S^{-}_{n}S^{+}_{n+1}\right)\Big]\, ,
\end{eqnarray}
where $S^{+}_{n}=S^{+}_{x} \pm i S^{+}_{y}$.

Using the Jordan-Wigner transformations [\onlinecite{JW_1928}]
\begin{eqnarray} S_{n}^{+} &=& a^{\dagger}_{n} \exp\left(i\pi\sum_{m < n}
a^{\dagger}_{m}a^{\phantom{\dagger}}_{m}\right)\, ,\\
\quad S_{n}^{-} &=&
\exp\left(-i\pi\sum_{m < n}a^{\dagger}_{m}a^{\phantom{\dagger}}_{m}\right)a_{n}\, ,\\
S^{z}_{n} &=& a^{\dagger}_{n}a^{\phantom{\dagger}}_{n} -
1/2,\label{jordanwigner} \end{eqnarray}
where $a^{\dagger}_{n}$ ($a^{\phantom{\dagger}}_{n}$) is a spinless
fermion creation (annihilation) operator on site $n$, we rewrite the
initial lattice spin Hamiltonian (\ref{Hamiltonian_XXZ_Alt_DM}) in
terms of interacting spinless fermions in the following way:
\begin{eqnarray}
\label{Hamiltonian_XXZ_Alt_DM_SF}{\cal H} &=&
\frac{J}{2}\sum_{n}\left(a^{\dagger}_{n}a^{\phantom{\dagger}}_{n+1}+
a^{\dagger}_{n+1}a^{\phantom{\dagger}}_{n}\right)\nonumber\\
&+&\frac{iD_{0}}{2}\sum_{n}\left(a^{\dagger}_{n}a^{\phantom{\dagger}}_{n+1}-
a^{\dagger}_{n+1}a^{\phantom{\dagger}}_{n}\right)\nonumber\\
&+&
\frac{iD_{1}}{2}\sum_{n}(-1)^{n}\left(a^{\dagger}_{n}a^{\phantom{\dagger}}_{n+1}-
a^{\dagger}_{n+1}a^{\phantom{\dagger}}_{n}\right)
\nonumber\\
&+&J_{z}\sum_{n}(a^{\dagger}_{n}a^{\phantom{\dagger}}_{n}-1/2)
(a^{\dagger}_{n+1}a^{\phantom{\dagger}}_{n+1}-1/2)\, .
\end{eqnarray}

We first discuss the exactly solvable $XX$ limit $J_{z}=0$. Indeed,
in absence of the Ising part of the spin exchange ($J_{z}=0$), the
Hamiltonian can be easily diagonalized in the momentum space.
Indeed, performing the Fourier transform,
\begin{eqnarray}
a_n &=&\frac{1}{\sqrt{L}}\sum_k \, a_k e^{ikn}\, ,
\end{eqnarray}
at $J_{z}=0$ we obtain
\begin{eqnarray}{\cal H} &=&\sum_{k} \left[\epsilon(k)
\,a^{\dagger}_{k}a^{\phantom{\dagger}}_{k} + i\Delta(k)
a^{\dagger}_{k}a^{\phantom{\dagger}}_{ k+\pi}\right]\,
,\label{H-JW_Fermions-MS} \end{eqnarray}
where
\begin{eqnarray}
\epsilon(k)&=& \left(J\cos k -D_{0}\sin
k\right)=J_{eff}\cos(k+q_{0})\, , \label{Epsilon-k}\\
\Delta(k)&=& D_{1}\cos k \label{Delta-k}
\end{eqnarray}
and
\begin{eqnarray}
J_{eff}&=& \sqrt{J^2+D_{0}^2}\label{J-eff}\\
q_{0}&=& \arctan(D_{0}/J)\, . \label{q-0}
\end{eqnarray}
Thus, in absence of the staggered component of the DM interaction
and ($D_{1}=0$) the excitation spectrum of the model is given by the
same dispersion relation as the standard $XX$ chain
\begin{eqnarray}\label{H_gapless}
{\cal H}_{0} &=&\sum_{k}
\epsilon(k)\,a^{\dagger}_{k}a^{\phantom{\dagger}}_{k}\, ,
\end{eqnarray}
but for a uniform shift $q_{0}$ in the momentum vector due to the
uniform part of the DM interaction [\onlinecite{Derzhko_1}]. The
system is characterized by two Fermi points $k_{F}^{\pm}
=\pm\frac{\pi}{2} - q_{0}$, so that in the ground state all states
with  $\pi/2 \leq |k+q_{0}|\leq \pi$ are occupied and those with
$|k+q_{0}|<\pi/2$ are empty.  The bandwidth is half filled, the
total magnetization of the system in the ground state as well as the
average value of the on-site spin vanishes
\begin{equation}
m=\frac{1}{L}\sum_{n}\la 0 |S_{n}^{z}|0 \ra=0\, .
\end{equation}
The vacuum spin current, determined via the chirality order
parameter [\onlinecite{Chubukov_91,Furusaki_08}] is evaluated as
\begin{eqnarray}
&&J_{sp}=\frac{1}{L}\sum_{n}\la 0| \kappa^{z}_{n}|0
\ra=\frac{2}{\pi}\sin q_{0}\, .
\end{eqnarray}
Note that due to the gapless excitation spectrum, all corresponding
correlations decay in power-laws [\onlinecite{Takahashi_book_99}]
and no LRO is present in absence of modulated part of the DM
interaction.

\begin{figure}[hb]
\begin{center}
\includegraphics[width =6.0cm,angle=0]{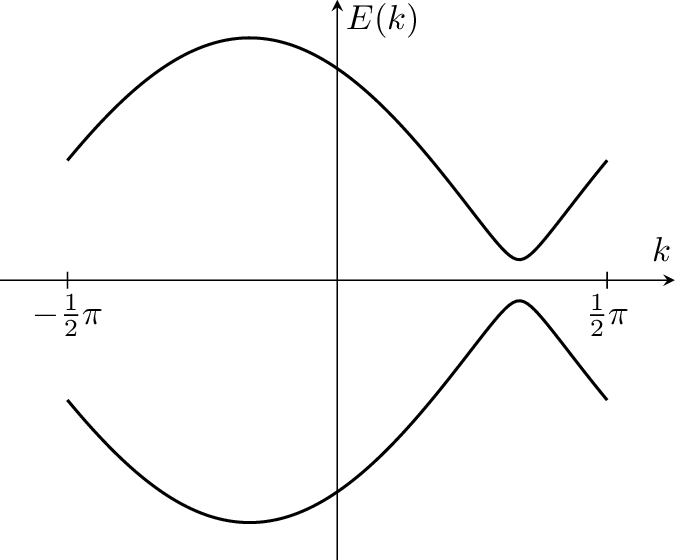}
\caption{Free spinless fermion (spinon) dispersion relation in the
case of finite uniform and alternating DM interaction. Here $J=1$,
$J_z=0$, $D_{0}=\tan(\pi/6)$ and $D_{1}=0.2$}\label{fig:Fig2}
\end{center}
\end{figure}

At $D_{1}\neq 0$, diagonalization of the Hamiltonian
(\ref{H-JW_Fermions-MS}) is also straightforward. It is convenient
to restrict momenta within the reduced Brillouin zone $-\pi/2 < k
\leq \pi/2$ and to introduce a new notation $a_{k+\pi}=b_{k}$. In
these terms the Hamiltonian reads
\begin{eqnarray}
&{\cal H} =\sum_{k}{}^{\prime} \left[\epsilon(k)\left(a_{k}^{\dag}
a^{\phantom{\dagger}}_{k}
-b_{k}^{\dag}b^{\phantom{\dagger}}_{k}\right)+i\Delta(k)\left(a_{k}^{\dag}b^{\phantom{\dagger}}_{k}
- b_{k}^{\dag}a^{\phantom{\dagger}}_{k}\right)\right]\, ,&\nonumber
\end{eqnarray}
where prime in the sum means that the summation is taken over the
reduced Brilluoin zone $-\pi/2 < k \leq \pi/2$. Using the unitary
transformation
\begin{eqnarray}
a_{k} &=& \cos\phi_{k}\,\alpha_{k} +i\sin\phi_{k}\,\beta_{k}\,
,\\
b_{k}&=& i\sin\phi_{k}\,\alpha_{k} +\cos\phi_{k}\,\beta_{k}\, .
\end{eqnarray}
and choosing
$$
\tan\left(2\phi_{k}\right)=-\Delta(k)/\epsilon(k)
$$
we obtain
\begin{eqnarray}
{\cal H}
&=&\sum_{\pi/2<k\leq\pi/2}E(k)\left(\alpha_{k}^{\dag}\alpha^{\phantom{\dagger}}_{k}
- \beta_{k}^{\dag}\beta^{\phantom{\dagger}}_{k}\right)
\end{eqnarray}
where
\bea\label{EigenValues-Gen}
E(k)&=&\sqrt{J^{2}_{eff}\cos^{2}(k+q_{0})
 + D_{1}^{2}\cos^{2} k} \eea
Note that  in absence of the uniform component of the DM interaction
($D_{0}=0,\, D_{1}\neq 0$), $E(k)=\pm \sqrt{J^{2} + D_{1}^{2}}\,\cos
k$ and therefore the excitation spectrum is gapless, the vacuum spin
current $J_{sp}=0$ and no LRO is present in the ground state.

Only at $D \neq 0,\, D_{1}\neq 0$ the spectrum is characterized by a
finite excitation gap (see Fig.\ \ref{fig:Fig2}).
\begin{eqnarray}\label{Gap_XX_D1}
\Delta_{exc}&=&J^{\ast}\sqrt{2\left(1-\sqrt{1-\left(2D_{0}D_{1}/J^{\ast}\right)^{2}}\right)}\nonumber\\
&&\hspace{8mm}\simeq 2D_{0}\,D_{1}/J^{\ast}\, ,
\end{eqnarray}
where
$$
J^{\ast}=\sqrt{J^{2}+D^{2}_{0}+D^{2}_{1}}\, .
$$

In the ground state of the gapped phase, all states in the negative
energy $\beta$-band are filled
$n_{\beta}(k)=\langle\beta^{\dagger}_{k}\beta^{\phantom{\dagger}}_{k}\rangle=1$,
while all the states in the positive energy $\alpha$-band are empty,
$n_{\alpha}(k)=\langle\alpha^{\dagger}_{k}\alpha^{\phantom{\dagger}}_{k}\rangle=1$.
As the result, in the ground state the $z$-projection of the total
spin
\begin{eqnarray}
M=\sum_{n}\la 0|S_{n}^{z}|0\ra =\frac{L}{2\pi}\int_{-\pi/2}^{\pi/2}
[\,n_{\beta}(k)-1/2\,]=0\,
\end{eqnarray}
as well as the staggered part of the on-site magnetization
\begin{eqnarray}
m=\frac{1}{L}\sum_{n}(-1)^{n}\la 0|S_{n}^{z}|0\ra=0\, .
\end{eqnarray}

It is straightforward to obtain, that the ground state average of
the staggered transverse spin dimerization and chirality order
parameters [\onlinecite{Chubukov_91,Furusaki_08}] are given by
\begin{eqnarray}
\epsilon_{\perp}&=&\frac{1}{L}\sum_{n}\,(-1)^{n}\la
0|\left(S^{+}_{n}S^{-}_{n+1}+S^{-}_{n}S^{+}_{n+1}\right)|0\ra
=\nonumber\\
&=& -\frac{D_{1}}{\pi}\int_{-\pi/2}^{\pi/2}\,\frac{\sin k \cos
k}{E(k)}\,dk \,
\end{eqnarray}
and
\begin{eqnarray}
\kappa&=&\frac{i}{L}\sum_{n}\,(-1)^{n}\la
0|\left(S^{+}_{n}S^{-}_{n+1}-S^{-}_{n}S^{+}_{n+1}\right)|0\ra
=\nonumber\\
&=&\frac{D_{1}}{\pi}\int_{-\pi/2}^{\pi/2}\,\frac{\cos^{2}k}{E(k)}\,dk
\,.
\end{eqnarray}
respectively.

It is easy to check by inspection, that both link-located order
parameters $\epsilon_{\perp} \rightarrow 0$ and $\kappa \rightarrow
0$ at $D_{0}=0$ and $D_{1} \neq 0$.

\begin{figure}[hb]
\begin{center}
\includegraphics[width =6.5cm,angle=0]{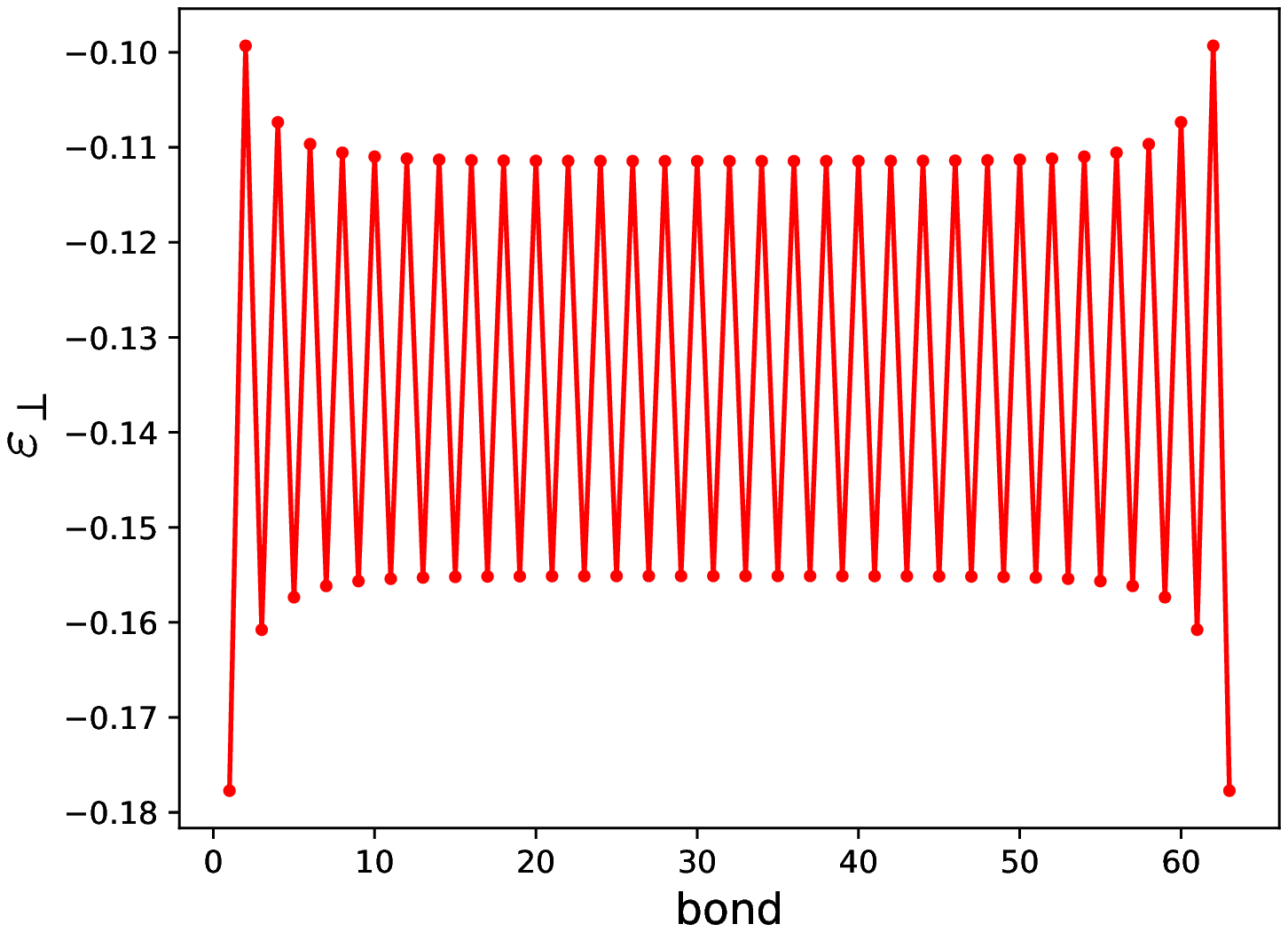}
\includegraphics[width =6.5cm,angle=0]{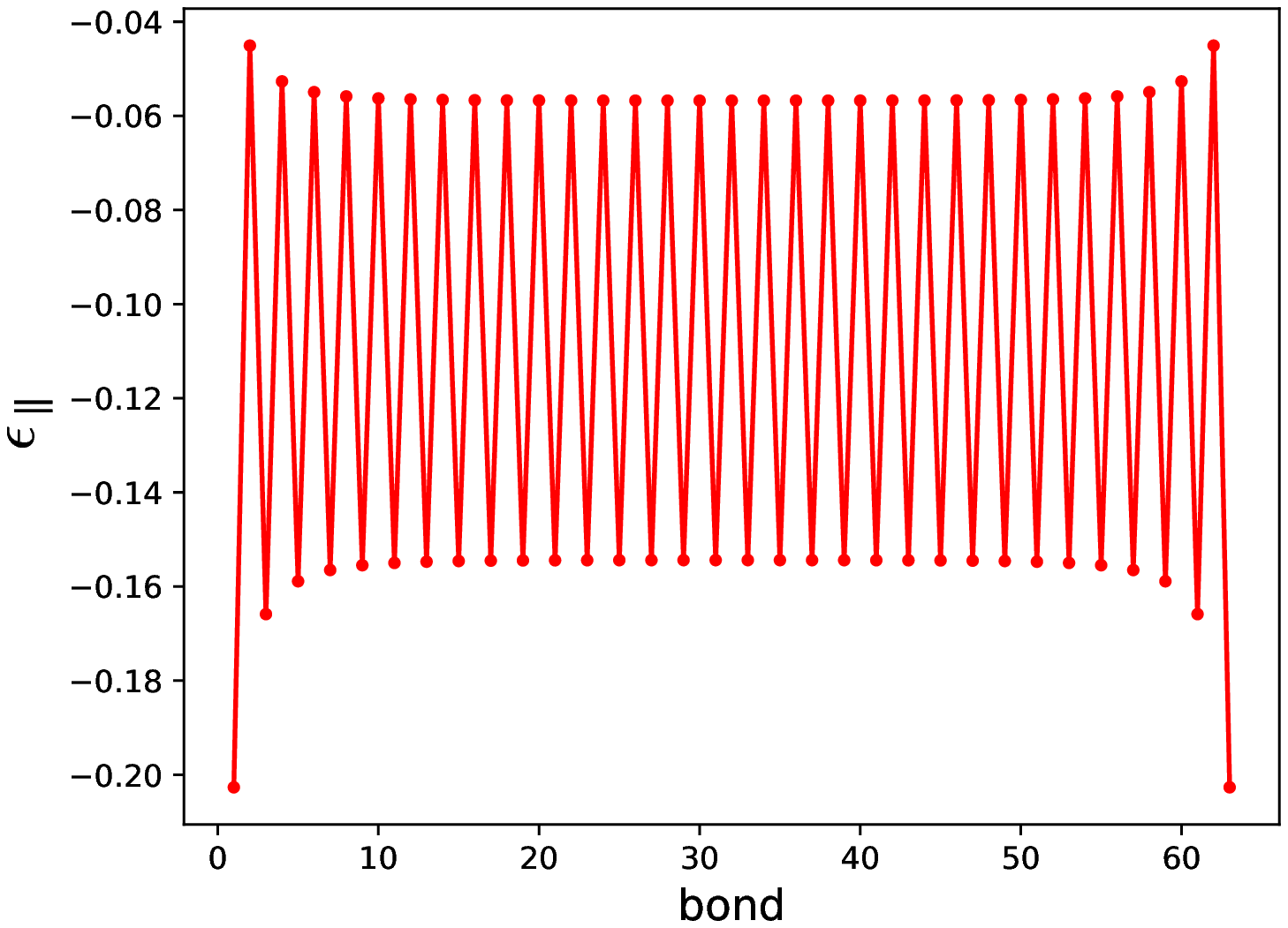}
\includegraphics[width =6.5cm,angle=0]{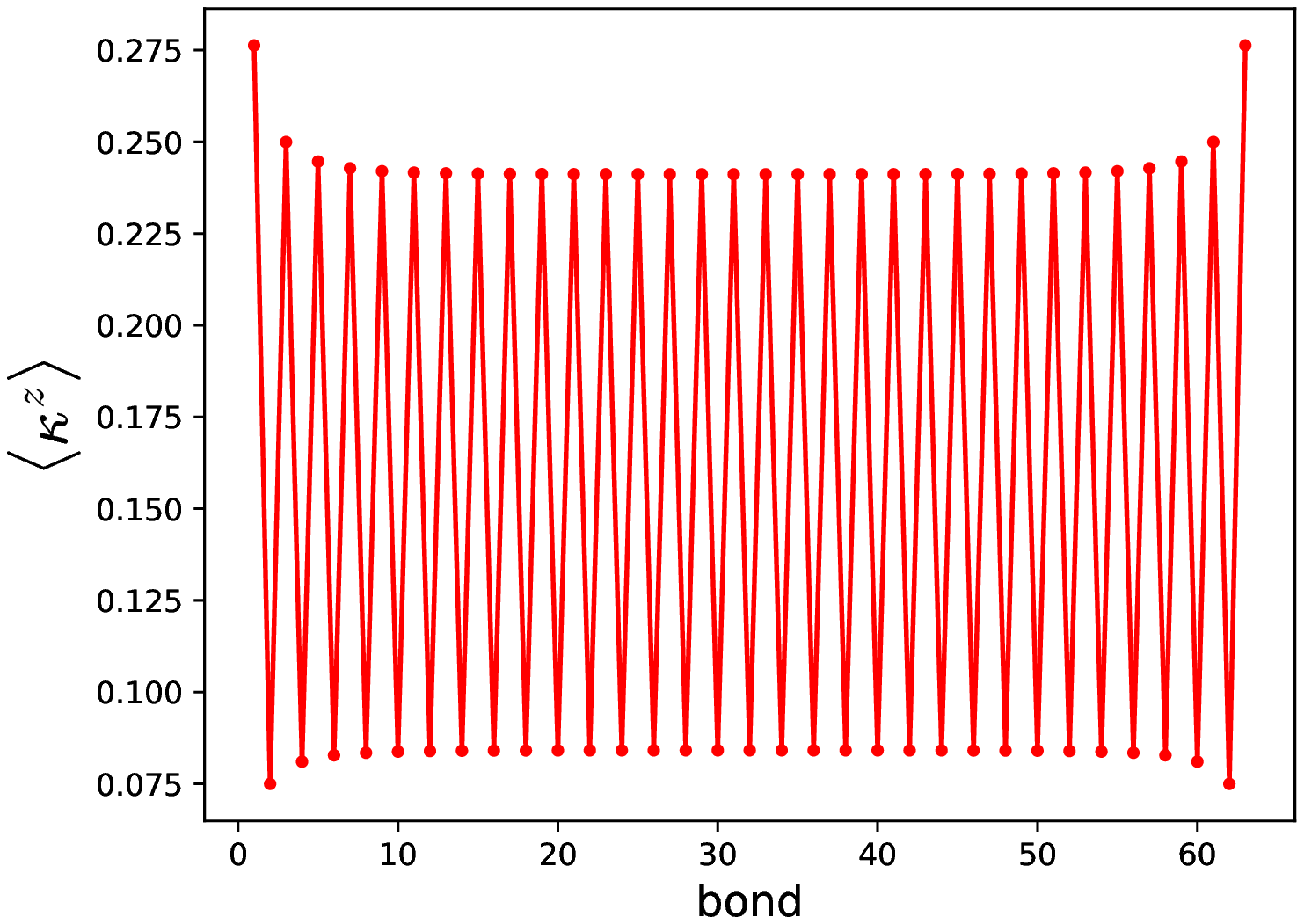}
\caption{
    The ground state expectation value distribution along bonds of: the
    transverse part of the nearest-neighbor spin-spin
    exchange operator $\epsilon_{n}^{\perp}$ (top panel),
    the longitudinal  part of the same operator $\epsilon_{n}^{\parallel}$ (middle panel), and
    the spin chirality operator $\kappa^z_{n}$ (bottom panel)
    in the case of finite uniform and alternating DM interactions.
    The results correspond to a chain of $L=64$ sites with OBC, with parameters $J=1$, $D_0=\tan(\pi/6)$ and $D_{1}=0.2$.
}\label{fig:Fig3}
\end{center}
\end{figure}

To conclude the considerations on the $XX$ limit of the model
(\ref{Hamilt_XXZ}), we present exact results of the local ground
state expectation values of the considered order parameters, aiming
to illustrate the described ordered patterns. They have been
obtained for finite chains of length $L=64$ with open boundary
conditions (OBC), also providing an insight into boundary features
found in DMRG computations for the interacting case (see Section
\ref{sec:numerics}).

In Fig.\ \ref{fig:Fig3} we have plotted the ground state distribution
of the {\em transverse} and {\em longitudinal} components of the
spin-exchange
\begin{eqnarray}
\epsilon_{\perp}(n)&=&\frac{1}{2}\la\left(S^{x}_{n}S^{x}_{n+1}+S^{y}_{n}S^{y}_{n+1}\right)\ra\,
,\\
\epsilon_{\parallel}(n)&=& \la\,S^{z}_{n}S^{z}_{n+1}\,\ra \, ,
\end{eqnarray}
and that of the $z$-component of the spin chirality vector, $\kappa^z_n$.
These show a well pronounced
alternating pattern in complete agreement with analytical results.
Notice that distortions close to the edges are a byproduct of OBC, thus in order
to compute bulk averages one usually discards a convenient number of sites at each boundary.
\begin{figure}[hb]
\begin{center}
\includegraphics[width =6.5cm,angle=0]{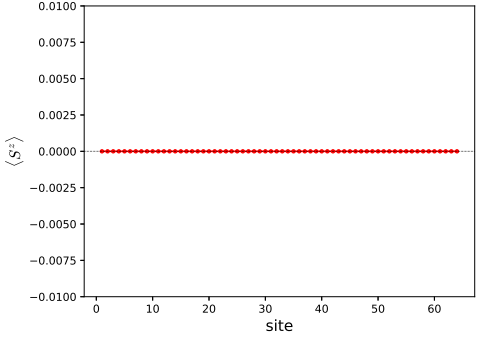}
\caption{The ground state expectation value distribution of the $z$
component of the spin operator as a function of site number. System
parameters are the same as in Fig.\ \ref{fig:Fig3}. }
\label{fig:Fig4}
\end{center}
\end{figure}
In Fig.\ \ref{fig:Fig4} we show the ground state distribution of
the on-site magnetization. In spite of the modulated terms in the
Hamiltonian one observes that the $z$-component of the spin density
is homogeneous and strongly zero, in marked contrast with the ground
state averages of the link-located order parameters.

\section{Gauging away the DM interaction}

To make next step forward, it is useful to rewrite the Hamiltonian
(\ref{Hamiltonian_XXZ_Alt_DM}) in a physically more suggestive
manner by rotating spins and gauging away the DM interaction term.
Here we follow the route, developed in the Ref.
[\onlinecite{Tsvelick_01}], in the case of a chain with uniform DM
interaction.

In the considered case of the Heisenberg chain with alternating DM
interaction, as a first step it is convenient to rewrite the
Hamiltonian in a way which explicitly takes into account doubling of
the unit cell by the staggered part of the DM interaction. Defining
new dimensionless parameters $d_{\pm}=(D_{0} \pm D_{1})/J$  the
Hamiltonian (\ref{Hamilt_XXZ}) reads
\begin{eqnarray}
\label{Hamilton_XXZ_w_Alt_DMI_2} {\cal H} &=&\frac{J}{2}
\sum_{m=1}^{N/2}\Big[\,\left(S^{+}_{2m-1}S^{-}_{2m} +
S^{-}_{2m-1}S^{+}_{2m}\right)\nonumber\\
&+&i\,d_{-}\left(S^{+}_{2m-1}S^{-}_{2m}
- S^{-}_{2m-1}S^{+}_{2m}\right)\nonumber\\
&&\hspace{5mm}+\left(S^{+}_{2m}S^{-}_{2m+1} +
S^{-}_{2m}S^{+}_{2m+1}\right)\nonumber\\
&+& i\,d_{+}\left(S^{+}_{2m}S^{-}_{2m+1} -
S^{-}_{2m}S^{+}_{2m+1}\right)\nonumber\\
&& \hspace{5mm}+2\,\Delta\, S^{z}_{2m}\left(S^{z}_{2m-1}
+S^{z}_{2m+1}\right)\Big] \, .
\end{eqnarray}
We introduce new spin variables ${\bf \tau}_{2m}$ and ${\bf
\tau}_{2m+1}$ by performing a site-dependent rotation of spins along
the chain around the $z$ axis with relative angle $\vartheta_{-}$
for spins at consecutive odd-even sites ($2m-1,2m$) and
$\vartheta_{+}$ for spins at consecutive even-odd sites ($2m,2m+1$),
so that
\begin{eqnarray}
&&  S^{+}_{2m-1}= e^{i(m-1)(\vartheta_{-}+\vartheta_{+})}\,\,
\tau^+_{2m-1}\, ,\nonumber\\
&& S^{+}_{2m} =  e^{im\vartheta_{-}+i(m-1)\vartheta_{+}}\,\,\tau^+_{2m}\, ,\label{transfo_1}\\
&& S^{+}_{2m+1}= e^{im(\vartheta_{-}+\vartheta_{+})}\,\,
\tau^+_{2m+1}\, ,\nonumber\\
&& S^z_{2m \pm 1}= \tau^z_{2m \pm 1}\,\qquad S^z_{2m}= \tau^z_{2m}\
.\nonumber 
\end{eqnarray}
In the new variables we obtain
\begin{eqnarray}
S_{2m-1}^{+}S_{2m}^{-}\,\pm h.c. &=&
\cos\vartheta_{-}\left(\tau_{2m-1}^{+}\tau_{2m}^{-}\, \pm \,
\tau_{2m-1}^{-}\tau_{2m}^{+}\right)\nonumber\\
&&\hspace{-15mm}- i
\sin\vartheta_{-}\left(\tau_{2m-1}^{+}\tau_{2m}^{-}\, \mp\,
\tau_{2m-1}^{-}\tau_{2m}^{+}\right),\label{Tau-Mapping_1a}
\end{eqnarray}
\begin{eqnarray}
S_{2m}^{+}S_{2m+1}^{-}\,\mp \, h.c.
&=&\cos\vartheta_{+}\left(\tau_{2m}^{+}\tau_{2m+1}^{-}\,
\pm \, \tau_{2m}^{-}\tau_{2m_1}^{+}\right)\nonumber\\
 &&\hspace{-15mm}- i
\sin\vartheta_{+}\left(\tau_{2m}^{+}\tau_{2m+1}^{-}\, \mp\,
\tau_{2m}^{-}\tau_{2m+1}^{+}\right).\label{Tau-Mapping_1b}
\end{eqnarray}
Inserting (\ref{Tau-Mapping_1a})-(\ref{Tau-Mapping_1b}) in
(\ref{Hamilton_XXZ_w_Alt_DMI_2}) we map the initial Hamiltonian onto
\begin{eqnarray} \label{XXZ_w_Modulated_DMI-rotated} &&{\cal H} =\frac{J}{2}
\sum_{m=1}^{N/2}\Big[\,\left(\cos\vartheta_{-}+d_{-}\sin\vartheta_{-}\right)
\left(\tau_{2m-1}^{+}\tau_{2m}^{-}+\tau_{2m}^{-}\tau_{2m-1}^{+}\right) \nonumber\\
&&\hspace{+4mm} -
i\,\left(\sin\vartheta_{-}-d_{-}\cos\vartheta_{-}\right)\left(\tau_{2m-1}^{+}\tau_{2m}^{-}\,
-\, \tau_{2m}^{-}\tau_{2m-1}^{+}\right)\nonumber\\
&&\hspace{+4mm}
\,+\left(\cos\vartheta_{-}+d_{+}\sin\vartheta_{+}\right)\left(\tau_{2m}^{+}\tau_{2m+1}^{-}\,
+\,
\tau_{2m}^{-}\tau_{2m+1}^{+}\right)\nonumber\\
&&\hspace{+4mm}
-i\left(\sin\vartheta_{-}-d_{+}\cos\vartheta_{+}\right)
\left(\tau_{2m}^{+}\tau_{2m+1}^{-}\,
-\, \tau_{2m}^{-}\tau_{2m+1}^{+}\right)\nonumber\\
&&\hspace{+4mm} + 2\,\Delta\,
\tau_{2m}^{z}\,\left(\tau_{2m-1}^{z}+\tau_{2m+1}^{z} \right)\Big] \,
.
\end{eqnarray}
Choosing angles $\vartheta_{\pm}$ such that
$$\tan\vartheta_{\pm}= d_{\pm},$$
one can cancel DM like terms
\begin{eqnarray}
\label{Cos-Sin_Theta} &&J(\sin\vartheta_{\pm}  - d_{\pm}
\,\cos\theta_{\pm }) = 0\,,\nonumber\\
 &&\hspace{-2mm} J(\cos\vartheta_{\pm}  + d_{\pm}
\,\sin\theta_{\pm }) =J\sqrt{1+d_{\pm}^{2}}\,\equiv J_{\pm}
\end{eqnarray}
and obtain the Hamiltonian without the DM interaction but only with
the alternating transverse exchange interaction
[\onlinecite{Derzhko_2}]
\begin{eqnarray}
\label{XXZ_w_Modulated_DMI_Rotated}{\cal H} &=&
\sum_{m=1}^{N/2}\Big[\,\frac{J_{-}}{2}\left(\tau^{+}_{2m-1}\tau^{-}_{2m}
+
\tau^{-}_{2m-1}\tau^{+}_{2m}\right)\nonumber\\
&&\hspace{5mm}+\, \frac{J_{+}}{2}\left(\tau^{+}_{2m}\tau^{-}_{2m+1}
+ \tau^{-}_{2m}\tau^{+}_{2m+1}\right)\nonumber\\
&&\hspace{5mm}+\, J_{z} \,\tau^{z}_{2m}\left(
\tau^{z}_{2m-1}+\tau^{z}_{2m+1}\right)\Big]\, .
\end{eqnarray}
It is instructive to rewrite the Hamiltonian
(\ref{XXZ_w_Modulated_DMI_Rotated}) in the following, more common,
form
\begin{eqnarray}
\label{XXZ_w_Modulated_DMI_Rotated_v2} {\cal H}&=& \tilde{J}
\sum_{n}
\Big[\,\frac{1}{2}(1+(-1)^{n}\delta)\left(\tau^{+}_{n}\tau^{-}_{n+1}
+ \tau^{-}_{n}\tau^{+}_{n+1}\right)\nonumber\\
 &&\hspace{12mm} \,+ \gamma^{\ast} \tau^{z}_{n}\tau^{z}_{n+1}\Big]\, ,
\end{eqnarray}
where, at $d_{i} \ll 1$ ( $i=\pm$),
\begin{eqnarray}
&&\hspace{-6mm}  \tilde{J}
=\frac{1}{2}\left(J_{+}+J_{-}\right)\simeq
J^{\ast}+{\cal O}\left(d_{i}^{4}\right),\\
&&\hspace{-6mm} \delta\, \tilde{J} =
\frac{1}{2}\left(J_{+}-J_{-}\right)\simeq
\frac{D_{0}D_{1}}{J^{\ast}}+{\cal O}\left(d_{i}^{4}\right)\,
\end{eqnarray}
and
\begin{eqnarray}\label{gamma-asterix}
\gamma^{\ast} = J_{z}/\tilde{J}\simeq J_{z}/J^{\ast}+{\cal
O}\left(d_{i}^{4}\right)\, .
\end{eqnarray}
At $J_{-}\neq J_{+}$ the Hamiltonian
(\ref{XXZ_w_Modulated_DMI_Rotated_v2}) is recognized as a
Hamiltonian of the $XXZ$ chain with alternating transverse exchange.
Note that the alternation of the transverse exchange $\delta \neq 0$
only for finite $D_{1}\neq 0$ and $D_{0} \neq 0$. In the following
we will discard ${\cal O}\left(d_{i}^{4}\right)$ corrections.

In the case of uniform DM interaction ($D_{1}=0$) the gauge
transformation reduces to the consecutive rotation of spins along
the chain around the $z$ axis with respect to the nearest neighbor
on the same angle
$$
\theta=\arctan\left(D_{0}/J\right).
$$
Because in this limit $J_{+}=J_{-}$ i.e. $\delta =0$,  the effect of
the uniform DM interaction reduces to the renormalization of the
exchange anisotropy $\gamma \rightarrow \gamma^{\ast}$ and change of
the boundary conditions. Respectively the Heisenberg chain with
uniform DM interaction is equivalent to an XXZ chain with twisted
boundary conditions.  In particular, the excitation spectrum and the
bulk correlation functions of a spin-1/2 XXZ Heisenberg chain with
DM interaction can be obtained from that of the corresponding XXZ
chain
\begin{eqnarray}
\label{Effective-tau_hamiltonian_2} &&\hspace{-5mm}{\cal
H}=J^{\ast}\sum_{n=1}^{N}\Big[\,\frac{1}{2}\left(\tau^{+}_{n}\tau^{-}_{n+1}
+ \tau^{-}_{n}\tau^{+}_{n+1}\right)+  \gamma^{\ast}
\tau^{z}_{n}\tau^{z}_{n+1}\Big] ,
\end{eqnarray}
taking into account the shift in momentum induced by the mapping
(\ref{transfo_1}) and renormalization of the anisotropy parameter
[\onlinecite{Tsvelick_01}].

In the case of staggered DM interaction $D(n)=(-1)^{n}D_{1}$
$$
\vartheta_{+}=-\vartheta_{-}=\vartheta=\arctan\left(D_{1}/J\right)
$$
and the gauge transformation becomes global and corresponds to the
rotation of all spins on even sites around the $z$ axis on the same
angle $\theta$
\begin{eqnarray}
S^{+}_{2m}&=& e^{i\theta}\,\tau^+_{2m}, \, S^z_{2m}= \tau^z_{2m+1}\
, \label{transfo_3a}
\end{eqnarray}
while the spins on even sites remain untouched:
\begin{eqnarray}
S^{+}_{2m-1}&=&  \tau^+_{2m-1}, \, S^z_{2m}= \tau^z_{2m}
.\label{transfo_3b}
\end{eqnarray}
This gives again the Hamiltonian
(\ref{Effective-tau_hamiltonian_2}), but with transverse exchange
$$
J^{\ast}=\sqrt{J^{2} +D_{1}^{2}} \, .
$$

Thus the effect of staggered DM interaction reduces only to the
enhancement of the exchange anisotropy and to the renormalization of
the bandwidth without any influence on the character of the
spectrum. For a system with open boundary conditions there are no
further changes, except for the appearance of gapless topological
edge states [\onlinecite{SSH_79}] that we take apart in numerical
computations. The bulk correlation functions of a spin-1/2
Heisenberg chain with staggered DM interaction can be obtained from
that of the corresponding XXZ chain
(\ref{Effective-tau_hamiltonian_2}) by taking into account the shift
on the relative angle $\theta$ between spins located on even and odd
sites.

The next step is to incorporate the effect of longitudinal part of
the spin exchange. Below we use the continuum-limit bosonization
approach to study low-energy properties of the Hamiltonian
(\ref{XXZ_w_Modulated_DMI_Rotated_v2}).

\section{The continuum-limit bosonization approach}

The continuum-limit bosonization approach to spin chains is well
known and discussed in detail in many excellent reviews and books.
Therefore, below we briefly sketch the most relevant steps and
bosonization conventions, while for technical details we refer the
reader to the corresponding references
[\onlinecite{Affleck_LN,GNT_Book,Giamarchi_Book}].

To obtain the continuum version of the Hamiltonian
(\ref{Effective-tau_hamiltonian_2}), we use the standard
bosonization expression of the spin operators
[\onlinecite{GNT_Book}]
\begin{eqnarray}
\tau_{n}^{z} &\simeq&  \sqrt{\frac{K}{\pi}} \partial_x \phi (x)+
(-1)^n
 \frac{{\it a}}{\pi \alpha} \sin\sqrt{4\pi K}\phi (x) \, ,\label{Tau-z_bos}\\
 \tau^{\pm}_{n} &\simeq& \frac{{\it b}}{\pi\alpha}
\cos(\sqrt{4\pi K}\phi)\,e^{\pm i \sqrt{\pi/ K}
\theta}\nonumber\\
& -&(-1)^{n}\frac{{\it c}}{\pi\alpha}\,e^{\pm i\sqrt{\pi/K}\theta}\,
.\label{Tau-pm_bos}
\end{eqnarray}
Here $\phi(x)$ and $\theta(x)$ are dual bosonic fields, $\partial_t
\phi =  u\partial_x \theta $, and satisfy the following
commutational relation
\begin{eqnarray}
\label{regcom}
&& [\phi(x),\theta(y)]  = i\Theta (y-x)\,,  \nonumber\\
&& [\phi(x),\theta(x)]  =i/2\, .
\end{eqnarray}
Here the non-universal real constants {\it a}, {\it b} and {\it c}
depend smoothly on the parameter $\gamma^{\ast}$, are of the order
of unity at $\gamma^{\ast}=0$
[\onlinecite{Hikihara_Furusaki_98,Lukyanov_Terras_03}] and are
expected to be nonzero everywhere at $|\gamma^{\ast}| < 1$. The
Luttinger liquid parameter is known within the critical line $ -1 <
\gamma^{\ast} < 1$ to be [\onlinecite{LP_75}]
\bea K &=& \frac{\pi}{2\arccos\left(-\gamma^{\ast}\right)}\, .
\label{K} \eea
Thus the parameter $K$ decreases monotonically from its maximal
value $K \to \infty$ at $\gamma^{\ast} \to -1$ (ferromagnetic
instability point), is equal to unity at $\gamma^{\ast} =0$
($J_{z}=0$) and reaches the value $K=1/2$ at $\gamma^{\ast} =1$
(isotropic antiferromagnetic chain). In the case of dominating Ising
type anisotropy, at $\gamma^{\ast} >1$, $K<1/2$.

Using (\ref{Tau-z_bos})-(\ref{Tau-pm_bos}) we finally obtain for
 the initial lattice Hamiltonian (\ref{XXZ_w_Modulated_DMI_Rotated}):
\begin{eqnarray}  \label{H_XXZ+D1_bos_DSG}
{\cal H}&=&u\int dx
\,\Big[\frac{1}{2}(\partial_{x}\phi)^2+\frac{1}{2}(\partial_{x}\theta)^2
+\frac{m_{0}}{\pi\alpha^{2}} \cos\sqrt{4\pi K}\phi \nonumber\\
&&\hspace{18mm}+\frac{M_{0}}{\pi\alpha^{2}}\cos\sqrt{16\pi K}\phi \,
\Big]\, ,
\end{eqnarray}
where
\begin{eqnarray}
m_{0} &\simeq& \delta = D_{0}D_{1}/J^{\ast\,2}\, ,\label{m}\\
M_{0} &\simeq& \gamma^{\ast}/2\pi\label{M}
\end{eqnarray}
and $u \simeq J^{\ast}/K$ stands for the velocity of spin
excitation.
Thus the effective continuum-limit version of the initial lattice
spin model (\ref{XXZ_w_Modulated_DMI_Rotated}) is given by the
double-frequency sine-Gordon (DSG) model [\onlinecite{DSG_Book_80}].
The DSG model (\ref{H_XXZ+D1_bos_DSG}) describes an interplay
between two perturbations to the Gaussian conformal field theory
with the ratio of their scaling dimensions equal to four. The DSG
model and its realizations in various 1D systems have been subject
of intensive studies in last decades
[\onlinecite{DSG_1,DSG_2a,DSG_2b,DSG_3a,DSG_3b,DSG_4,DSG_5,DSG_6,JN_19}].
It has been shown [\onlinecite{DSG_1}], that the ground state
properties of the DSG model are controlled by the scaling dimensions
of the  two {\em cosine} terms
$$
d=dim[\cos\sqrt{4\pi K}\phi]=K
$$
and
$$d^{\ast}=dim[\cos\sqrt{16\pi K}\phi]=4K\,
$$
present in the Hamiltonian.
Each of these {\em cosine} terms becomes relevant in the parameter
range where the corresponding scaling dimensionality $d \leq 2$ or
$d^{\ast}\leq 2$. Using (\ref{K}) we find that $d \leq 2$, i.e.\ the
first {\em cosine} term in (\ref{H_XXZ+D1_bos_DSG}) is relevant, at
$ \gamma^{\ast}
>\gamma^{\ast}_{c1}=-\sqrt{2}/2$, while $d^{\ast}\leq 2$, i.e.\ the second {\em
cosine} term in (\ref{H_XXZ+D1_bos_DSG}), for $\gamma^{\ast}> 1$.
This gives following four segments of the model parameter range (see
Fig.\ 1), where each one corresponds to the different mechanisms of
formation of the ground-state properties of the system:

\subsection{The Ferromagnetic sector $\gamma^{\ast} \leq -1$}
\label{subsec:A}

At $\gamma^{\ast} \leq -1$ the system is in the {\em ferromagnetic
phase}, all spins are oriented along the z-axis
$$\langle \tau_{n}^{z} \rangle = \langle S_{n}^{z} \rangle =1/2;\\
\langle \tau_{n}^{x}\rangle = \langle \tau_{n}^{y} \rangle = 0
$$
and therefore the effect of the DM interaction is completely
suppressed.

\subsection{The Luttinger-liquid sector $-1<\gamma^{\ast}< \gamma^{\ast}_{c1}$}
\label{subsec:B}

At  $-1<\gamma^{\ast}< \gamma^{\ast}_{c1}$, $d^{\ast} > d >2$ and
therefore both cosine terms in (\ref{H_XXZ+D1_bos_DSG}) are
irrelevant and can be neglected. The gapless long-wavelength
excitations of the anisotropic spin chain are described by the
standard Gaussian theory with the Hamiltonian
\begin{eqnarray}\label{SpinChainBosHam_Gauss}
{\cal H}_{0}& = &u\int dx \, \Big[\,\frac{1}{2}(\partial_x \phi)^{2}
 + \frac{1}{2}(\partial_x
\theta)^{2}\,\Big].
\end{eqnarray}
In this {\em critical Luttinger-liquid phase}, all correlations show
a power-law decay, with indices smoothly depending on the parameter
$K$ [\onlinecite{GNT_Book}].

\subsection{The dimerized sector $\gamma^{\ast}_{c1} < \gamma^{\ast} \leq 1$} 
\label{subsec:C1}

At $\gamma^{\ast}_{c1} < \gamma^{\ast} \leq 1$, $d<2$ while
$d^{\ast}>2$, therefore the double-frequency cosine term is
irrelevant and can be neglected. In this case infrared properties of
the system are described by the standard sine-Gordon (SG) model
\begin{eqnarray}  \label{H_XXZ+D1_bos_SG}
&\hspace{-3mm}{\cal H}=u\int dx
\,\Big[\frac{1}{2}(\partial_{x}\phi)^2+\frac{1}{2}(\partial_{x}\theta)^2
+\frac{m_{0}}{\pi\alpha^{2}} \cos\sqrt{4\pi K}\phi  \Big].&
\end{eqnarray}
With increasing $\gamma^{\ast}$, the scaling dimensionality of the
{\em relevant cosine} term changes from the marginal value $d=2$ at
$\gamma^{\ast}=\gamma^{\ast}_{c1}$, to $d=1/2$ at $\gamma^{\ast}=1$.
Thus, at $\gamma^{\ast}=\gamma^{\ast}_{c1} \simeq -0.7$ the BKT
[\onlinecite{KT_73}] quantum phase transition takes place in the
ground state of the system, the excitation gap opens at
$\gamma^{\ast}=\gamma^{\ast}_{c1}$ and remains finite in the whole
region $-0.7 < \gamma^{\ast} \leq 1$.

From the exact solution of the quantum sine-Gordon model
[\onlinecite{SG_exact_Sol,Zamolodchikov_95}] it is known that for
arbitrary finite $m_{0}$ the gapped excitation spectrum of the
Hamiltonian Eq.\ (\ref{H_XXZ+D1_bos_SG}) at $ 2 > d > 1$  ($-0.7 <
\gamma^{\ast}\leq 0$), consists of solitons and antisolitons with
masses
\begin{equation}  \label{Soliton_Mass}
{\cal M}_{sol} \sim
\left(m_{0}/J^{\ast}\right)^{\frac{1}{2-d}}=\left(m_{0}/J^{\ast}\right)^{\frac{1}{2-K}}
,
\end{equation}
while at $1 > d \geq 1/2$ ($0 < \gamma^{\ast}\leq 1$)  in addition,
also of soliton-antisoliton bound states ("breathers") with the
lowest breather mass
\begin{equation}  \label{Breathers_Mass}
{\cal M}_{br}=2{\cal M}_{sol}\sin\left(\frac{\pi K}{4-2K}\right)\, .
\end{equation}
Thus, in the whole parameter range $0 < \gamma^{\ast}\leq 1$ the
soliton mass ${\cal M}_{sol}$ is the energy scale which determines
the size of the spin excitation gap.

The excitation gap is exponentially small at the BKT phase
transition point
\begin{equation}  \label{Excitation_Gap_1}
\Delta_{exc}\sim J^{\ast}
\exp\left(-1/(\gamma^{\ast}-\gamma^{\ast}_{c1})\right)\, ,
\end{equation}
it smoothly increases with increasing $\gamma^{\ast}$, and at
$\gamma^{\ast}=0$
\begin{equation}  \label{Excitation_Gap_1}
\Delta_{exc}=2J^{\ast}{\cal M}_{sol}=2m_{0}=2D_{0}D_{1}/J^{\ast}\, ,
\end{equation}
in a perfect agreement with results obtained in the Sec.\
\ref{sec:TheModel} (see Eq.\ (\ref{Gap_XX_D1})). Finally, at
$\gamma^{\ast}=1$ the gap is
\begin{equation}
\label{Excitation_Gap_2} \Delta_{exc}=J^{\ast}{\cal M}_{br}
= J^{\ast}\left(
 D_{0}D_{1}/J^{\ast \,2}\right)^{2/3} \, .
\end{equation}

The gap in the excitation spectrum leads to suppression of
fluctuations in the system and the  $\phi$ field is condensed in one
of its vacua ensuring the minimum of the dominating potential energy
[\onlinecite{ME}]
\begin{eqnarray}  \label{Ordered-Field_1}
 \sqrt{4\pi K}\langle \phi \rangle  = \left\{
\begin{array}{l}
\pi \hskip0.5cm \textrm{at \quad $m_{0} > 0$} \\
0 \hskip0.5cm \textrm{ at \quad $m_{0} < 0$}
\end{array}
\right. \, .
\end{eqnarray}

As it follows from (\ref{Tau-z_bos})-(\ref{Tau-pm_bos}) trapping of
the $\phi$ field in one of the vacua from the set given by
(\ref{Ordered-Field_1}) leads to suppression of the site-located
magnetic degrees of freedom
$$\langle \tau_{n}^{z} \rangle = \langle \tau_{n}^{x}
\rangle = \langle \tau_{n}^{y} \rangle = 0.
$$
Respectively, using (\ref{transfo_1}) we obtain, that the
site-located magnetic order is also fully suppressed in the initial
spin chain system:
$$
\langle S_{n}^{z} \rangle = \langle S_{n}^{x} \rangle = \langle
S_{n}^{y} \rangle = 0.
$$

Moreover, if we consider the link-located degrees of freedom, using
(\ref{Tau-z_bos})-(\ref{Tau-pm_bos}) one obtains that the continuum
limit bosonized version of the $\tau$-spin chirality operator is
given by
\begin{eqnarray}
&&\kappa_{n}^{(\tau)}=-i\left(\tau^{+}_{n}\tau^{-}_{n+1}-h.c.\right)\rightarrow\nonumber\\
&&\hspace{4mm}\rightarrow\frac{2}{\sqrt{\pi}}\partial_{x}\theta
+(-1)^{n}\frac{{\it 2b}}{\pi\alpha}\sin(\sqrt{4\pi K}\phi)\,
\label{mapping_Kappa-Tau}
\end{eqnarray}
and therefore in the gapped phase, where $\sqrt{4\pi K}\la\phi\ra=0
\quad \, mod \quad \pi$
\begin{eqnarray}\label{Kappa_Tau_aver}
\langle \kappa_{n}^{(\tau)}\rangle=0.
\end{eqnarray}
However, the bosonized expressions for the staggered parts of the
$\tau$-spin longitudinal and transverse nearest-neighbor spin
exchange operators
\begin{eqnarray}
\epsilon_{\perp}^{(\tau)}(n)&=&\frac{(-1)^{n}}{2}\left(\tau^{+}_{n}\tau^{-}_{n+1}+h.c.\right)
\sim
\nonumber\\
& \sim& 
\frac{{\it a}}{2\pi^{2}\alpha^{2}}\cos(\sqrt{4\pi K}\phi)\label{mapping_Epsilon_Tau}\\
\epsilon_{z}^{(\tau)}(n)&=&(-1)^{n}\tau^{z}_{n}\tau^{z}_{n+1}
\sim\frac{{\it b}}{\pi\alpha}\cos(\sqrt{4\pi
K}\phi)\label{mapping_Tau-zz-staggered}
\end{eqnarray}
are characterized a finite vacuum expectation value in the gapped
phase and therefore, in the given gapped sector of the phase diagram
we find the presence of the long-range dimerization pattern in the
ground state:
\begin{eqnarray}
&&(-1)^{n}\langle\,\epsilon_{\perp}^{(\tau)}(n)\,\rangle \sim
(-1)^{n}\langle\,\epsilon_{z}^{(\tau)}(n)\,\rangle \simeq \epsilon
\label{Epsilon_Tau_z_aver}
\end{eqnarray}
where
\begin{eqnarray}
\epsilon = \langle\,\cos{\sqrt{2 \pi K}\phi}\,\rangle \simeq
m_{0}^{K}= \left(D_{0}D_{1}/J^{\ast \,2}\right)^{K}
\end{eqnarray}
at weak coupling ($m_{0}<< J^{\ast})$ and becomes of the unit order
in the strong coupling, at $m_{0} \geq J^{\ast}$
[\onlinecite{Luk_Zam_97}].

Using (\ref{Epsilon_Tau_z_aver}), from
(\ref{Tau-Mapping_1a})-(\ref{Tau-Mapping_1b}) we obtain, that in the
gapped phase the initial spin chain shows a long-range dimerization
order
\begin{eqnarray}
&& \hspace{-3mm}\frac{1}{L}\sum_{n}(-1)^{n}
\la\,\textbf{S}_{n}\cdot\textbf{S}_{n+1} \, \ra \sim
(\cos\vartheta_{+}-\cos\vartheta_{-})\,\epsilon \,,
\label{S-Dimerization-Pattern}
\end{eqnarray}
which coexists with the long-range order pattern of the alternating
spin chirality vector
\begin{eqnarray}
&& \hspace{-3mm}\frac{1}{L}\sum_{n}(-1)^{n}\la \,\kappa^{z}_{n}\ra
\sim
(\sin\vartheta_{+}-\sin\vartheta_{-})\,\epsilon \,.
\label{S-Chirality-Pattern}
\end{eqnarray}

\subsection{The Ising type sector $\gamma^{\ast}>1$}
\label{subsec:C2}

At $\gamma^{\ast}>1$ both cosine terms in (\ref{H_XXZ+D1_bos_DSG})
are relevant and, in principle, have to be considered on equal
grounds. Therefore in this case the low-energy sector of the initial
spin chain is given in terms of the double sine-Gordon model
\begin{eqnarray}  \label{H_XXZ+D1_bos_DSG_2}
{\cal H}&=&u\int dx
\,\Big[\frac{1}{2}(\partial_{x}\phi)^2+\frac{1}{2}(\partial_{x}\theta)^2
+\frac{m_{0}}{\pi\alpha^{2}} \cos\beta\phi \nonumber\\
&&\hspace{18mm}+\frac{M_{0}}{\pi\alpha^{2}}\cos2\beta\phi \, \Big]\,
,
\end{eqnarray}
with $\beta=\sqrt{4\pi K}$, which describes an interplay between two
relevant perturbations to the Gaussian conformal field theory
$H_{0}$ (\ref{SpinChainBosHam_Gauss}) with the ratio of their
scaling dimensions equal to 4. Since at $\gamma^{\ast}>1$ the
Luttinger parameter is $K<1/2$,  the parameter $\beta$ satisfies the
inequality $\beta^{2}=4\pi K < 4\pi$ and no extra relevant terms are
generated via the renormalization procedure. In consequence, the
description of the system is closed within the Hamiltonian (\ref
{H_XXZ+D1_bos_DSG_2}) [\onlinecite{DSG_1}]. Moreover, the very
presence of two {\em independent} model parameters, $\delta$ and
$\gamma^{\ast}$, which determine the bare values of masses of two
competing cosine terms, makes the phase diagram of the model rich
and opens the possibility to manipulate the low-energy properties of
the system by changing intensity of the DM interaction.

Since both terms are relevant, acting separately, each leads to the
pinning of the field $\phi$ in corresponding minima, however because
these two perturbations have different parity symmetries, the field
configurations which minimize one perturbation do not minimize the
other.

Indeed the vacuum expectation value $\la\,\phi\,\ra=\sqrt{\pi/16
K}$, which corresponds to the minimum of the $M_{0}\cos\sqrt{16\pi
K}\phi$ term, leads to the suppression of contributions coming from
the $m_{0}\cos\sqrt{4\pi K}\phi$ term, while trapping of the field
at the minima $\la\,\phi\,\ra=0, or \, \sqrt{\pi/4 K}$, which ensure
minimum of the latter {\em cosine} term, correspond to the maximum
of the former, double-frequency {\em cosine} potential. This
competition between possible sets of vacuum configurations of the
two {\em cosine} terms is resolved via the presence of the quantum
phase transition (QPT) in the ground state.

The very presence of the QPT can already be traced performing
minimization of the potential
\begin{eqnarray}\label{DSG_Potential}
{\cal V}(\phi)&=&m_{0}\cos\beta\phi + M_{0}\cos2\beta\phi\, ,
\end{eqnarray}
where the transition corresponds to the crossover from a double well
to a single well profile of the potential (\ref{DSG_Potential}).
indeed, one can easily obtain, that at $M_{0}>m_{0}/4$ the vacuum
expectation value of $\phi$ field which minimizes ${\cal V}(\phi)$
is given by (\ref{Ordered-Field_1}) and therefore in this case the
dimerized phase is realized ground state. However, at
$M_{0}>m_{0}/4$ the $\phi$ field is condensed in the minima
\begin{eqnarray}  \label{Ordered-Field_2}
 \,\langle \phi \,\rangle = \phi_{0}
=\frac{1}{\beta}\arccos\left(m_{0}/4M_{0}\right)
\end{eqnarray}
and, as the result, in addition to the dimerization pattern
\begin{eqnarray}  \label{Ordered-Field_2a}
&&(-1)^{n}\langle\,\epsilon_{i}^{(\tau)}(n)\,\rangle \sim
\langle\,\cos(\sqrt{4\pi K}\phi_{0})\rangle \,\, i=\perp,z
\end{eqnarray}
the ground state of the $\tau$-spin system is characterized by the
long range antiferromagnetic  order with the amplitude of the
staggered magnetization
\begin{eqnarray}  \label{Ordered-Field_2b}
&& m=(-1)^{n} \langle\,\tau^{z}_{n}\,\rangle \sim \sin\sqrt{4\pi
K}\phi_{0}\, .
\end{eqnarray}

Following the analysis, developed in the Ref.\ [\onlinecite{DSG_1}]
one can show that the model displays an Ising criticality with
central charge $c = 1/2$ on a quantum critical line. The critical
properties of this transition have been investigated in detail by
mapping the DSG model onto the deformed quantum Ashkin-Teller model
[\onlinecite{DSG_2b}]. The dimensional arguments based on equating
physical masses produced by the two cosine terms separately is
usually used to define the critical line. Using (\ref{Soliton_Mass})
one finds
\begin{equation}
\left\{ \begin{array}{lll}
m& = & m_{0}^{1/\left(2-K\right)}\\
M&=&   M_{0}^{1/\left(2-4K\right)}
\end{array}
\right.
\end{equation}
Equating these two masses we obtain the following expression for the
critical value of the chain anisotropy parameter vs. DM coupling:
\begin{equation}
\label{criticalline}
\gamma^{\ast}_{c2}=1+\left(\frac{D_{0}D_{1}}{J^{\ast \,
2}}\right)^{\frac{2-4K}{2-K}}\, .
\end{equation}
Because at $\gamma^{\ast}\gg 1$ the parameter $K$ has to approach
its minimal value $K \simeq 1/4$, we take as the transition point $K
\sim 1/3$ and therefore from (\ref{criticalline}) we obtain the
following rather rough estimate for the critical value of the
longitudinal exchange
$$J^{z}_{c} \sim 1+
D_{0}D_{1}/J^{\ast}.
$$

Below the critical point the system is in the dimerized phase, while
$\gamma^{\ast}>\gamma^{\ast}_{c2}$ the field is condensed in a such
vacuum minima $\langle \phi \,\rangle = \phi_{0}$ where $\sqrt{4\pi
K}\phi_{0} \neq 0,\pi/2$. Therefore, in this case the composite
ordered phase with coexisting {\em dimer} and {\em
antiferromagnetic} order is realized in the ground state of the
$\tau$-spin chain.

It is evident that in this phase the initial spin-chain, besides the
dimerization and chirality order, given by
(\ref{S-Dimerization-Pattern})-(\ref{S-Chirality-Pattern}), shows a
long range antiferromagnetic arrangement of the $z$-projections of
the ${\bf S}$-spins.

\section{Numerical Results}\label{sec:numerics}

In order to investigate the detailed behavior of the ground state
phase diagram and to test the validity of the picture obtained from
the continuum bosonization treatment, we present in this Section
results of numerical calculations for finite chains with open
boundary conditions, obtained with the DMRG technique
[\onlinecite{White_92}].

\begin{figure}[ht!]
\begin{center}
\includegraphics[width =7.5cm,angle=0]{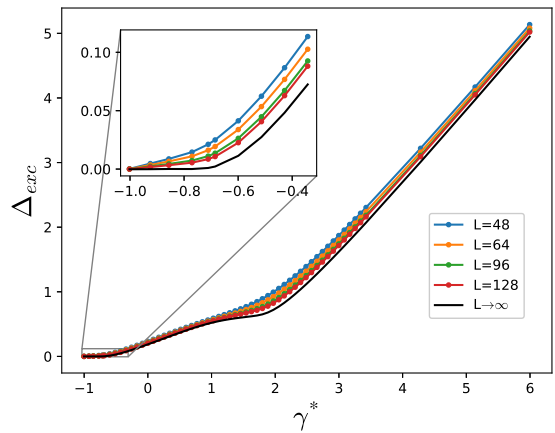}
\caption{ Spin excitation gap above the $S^z_{\rm total}=0$ ground
state. In the inset one can see a gapless phase. DMRG results for
finite chains of lengths $L = 48, 64, 96, 128$ and their $1/L$
extrapolations (black line) for a wide range $\gamma^{\ast}>-1$. }
\label{fig:DMRG-gap}
\end{center}
\end{figure}

 The computations were carried out for
finite-length systems with $L = 48, 64, 96$ and $128$ sites, using
the ALPS library [\onlinecite{ALPS_1,ALPS_2}].  System parameters
are set to  $J=1$, $D_{0} = tan(\pi/6)$ and $D_{1} = 0.2$, while the
bare value of the anisotropy $\Delta$ is varied providing values of
$-1 < \gamma^{\ast} \leq 6$. This restricts the ground state
analysis to the $S_z^{tot} = 0$ subspace. Keeping m = 400 states and
performing 10 sweeps we reproduced exact energies and expectation
values at $\gamma^{\ast}= 0$ for the same lengths with accuracy of
at least 6 digits.
Open boundary conditions on the alternating coupling have been
chosen in a topologically trivial sector, so as to avoid gapless
edge states. Averages of local expectation values are computed in
the central half of each chain in order to minimize open boundary
effects.

In Fig.\ \ref{fig:DMRG-gap} we show the excitation gap numerically
computed as
\begin{equation}
\label{Excitation_Gap_DMRG} \Delta_{exc}= E_{0}(N+1)+E(N+1)-2E(N)\,
,
\end{equation}
where $E_{0}(f)$ is the lowest energy state in the subspace with
fermionic occupation number $f$ and $N=L/2$ corresponds to the
$S_{z}^{tot} = 0$ subspace. In the inset one can appreciate the
gapless LL region described in subsection \ref{subsec:B}, in full
agreement with bosonization predictions. It is also apparent the
exponentially small gap opening at  $\gamma^{\ast}_{c1}\sim 0.7$,
supporting the presence of the BKT transition discussed in
subsection \ref{subsec:C1}. The gap value for $\gamma^{\ast} = 0$ of
course coincides with exact results in section \ref{sec:TheModel}.
\begin{widetext}

\begin{figure}[ht!]
\begin{center}
\includegraphics[width =7.5cm,angle=0]{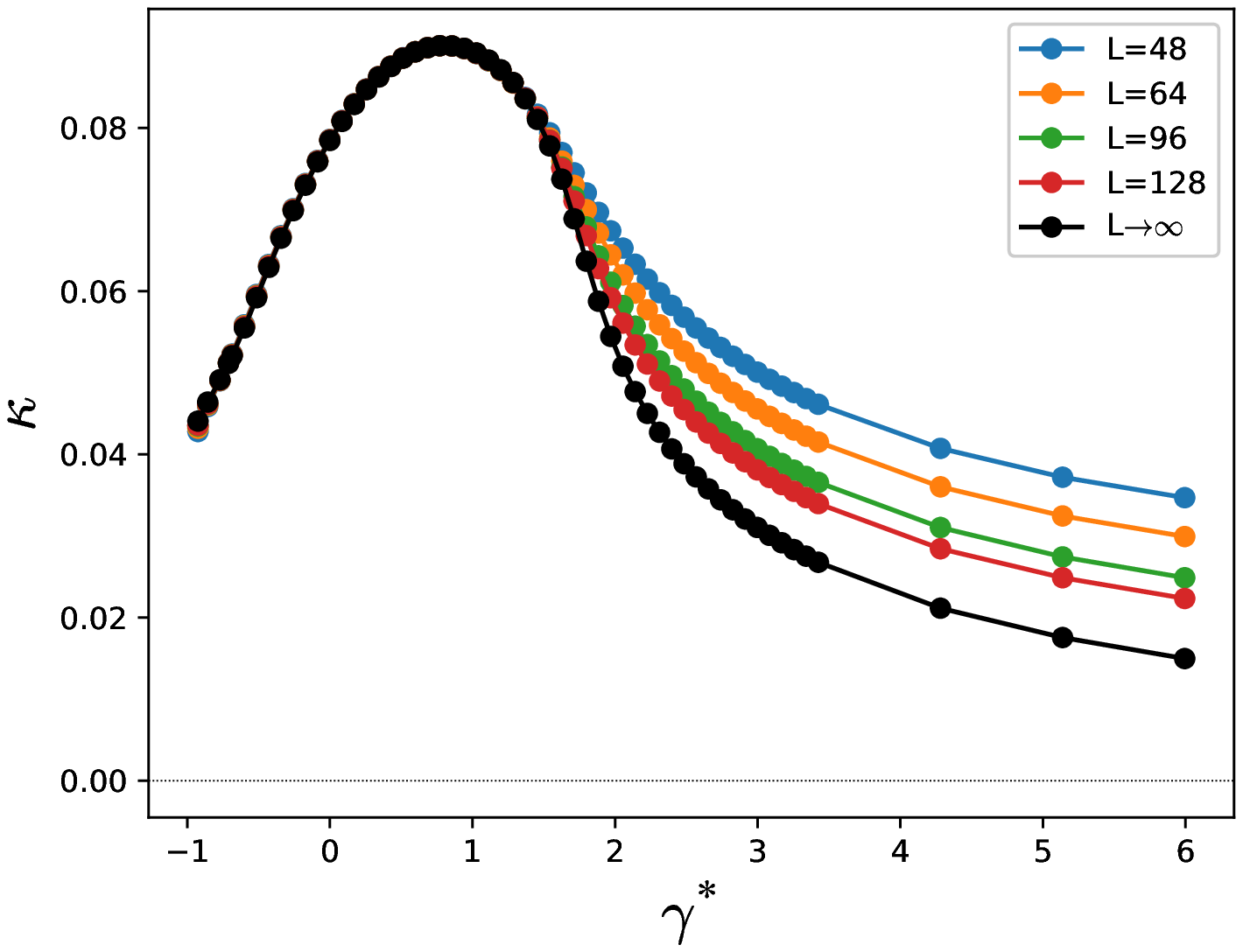} \hspace{1.6cm}
\includegraphics[width =7.5cm,angle=0]{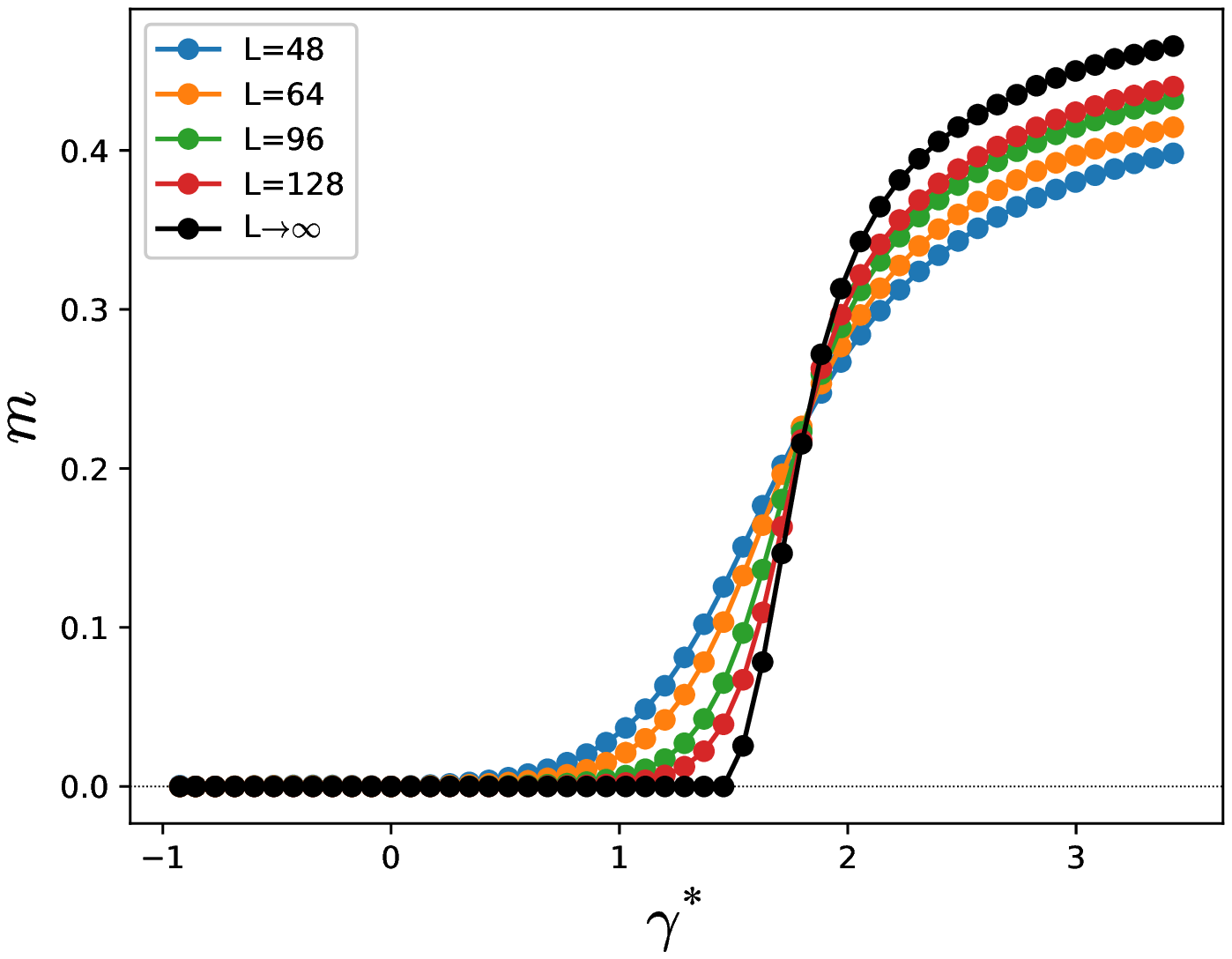}
\includegraphics[width =7.5cm,angle=0]{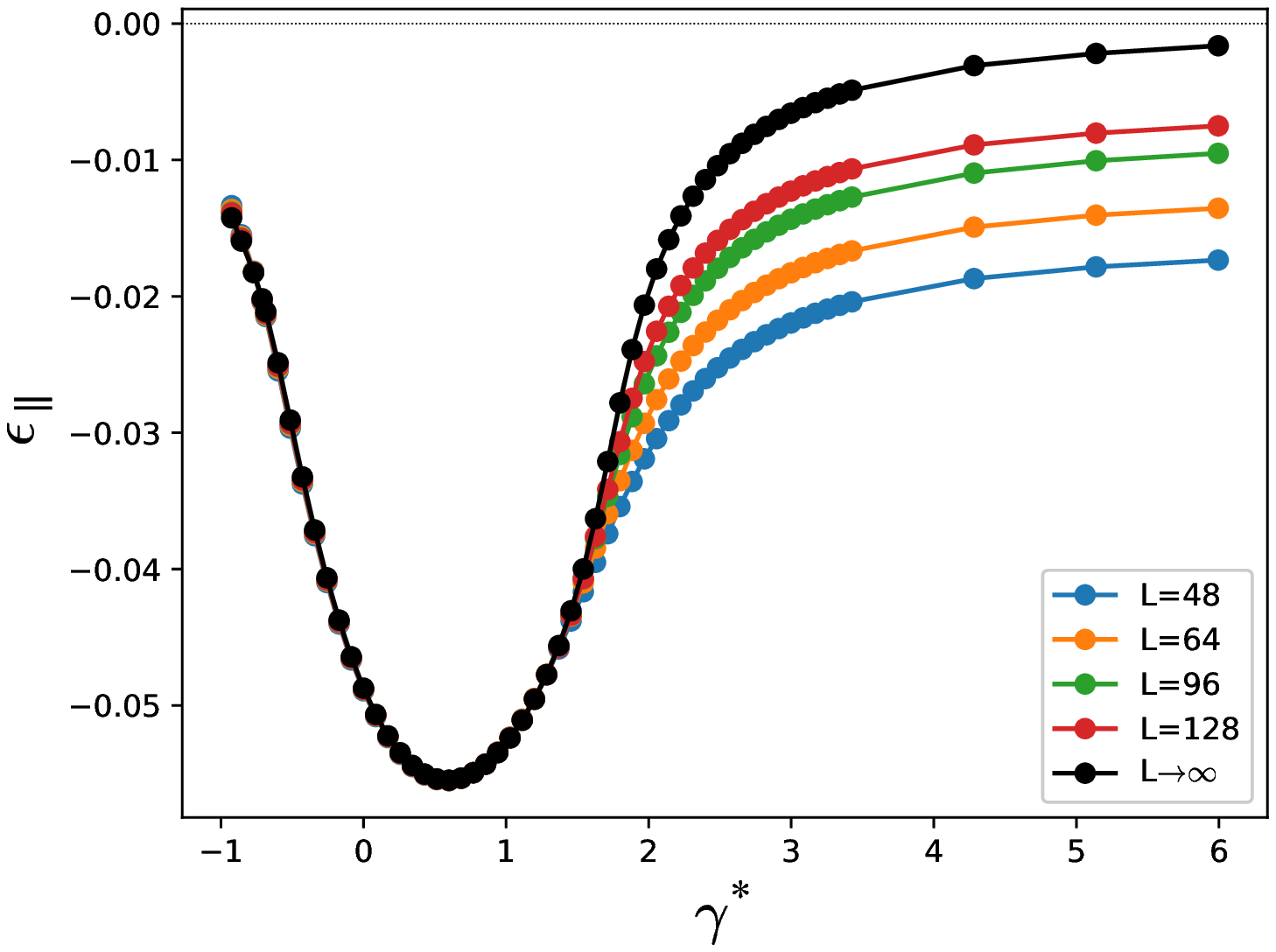} \hspace{1.6cm}
\includegraphics[width =7.5cm,angle=0]{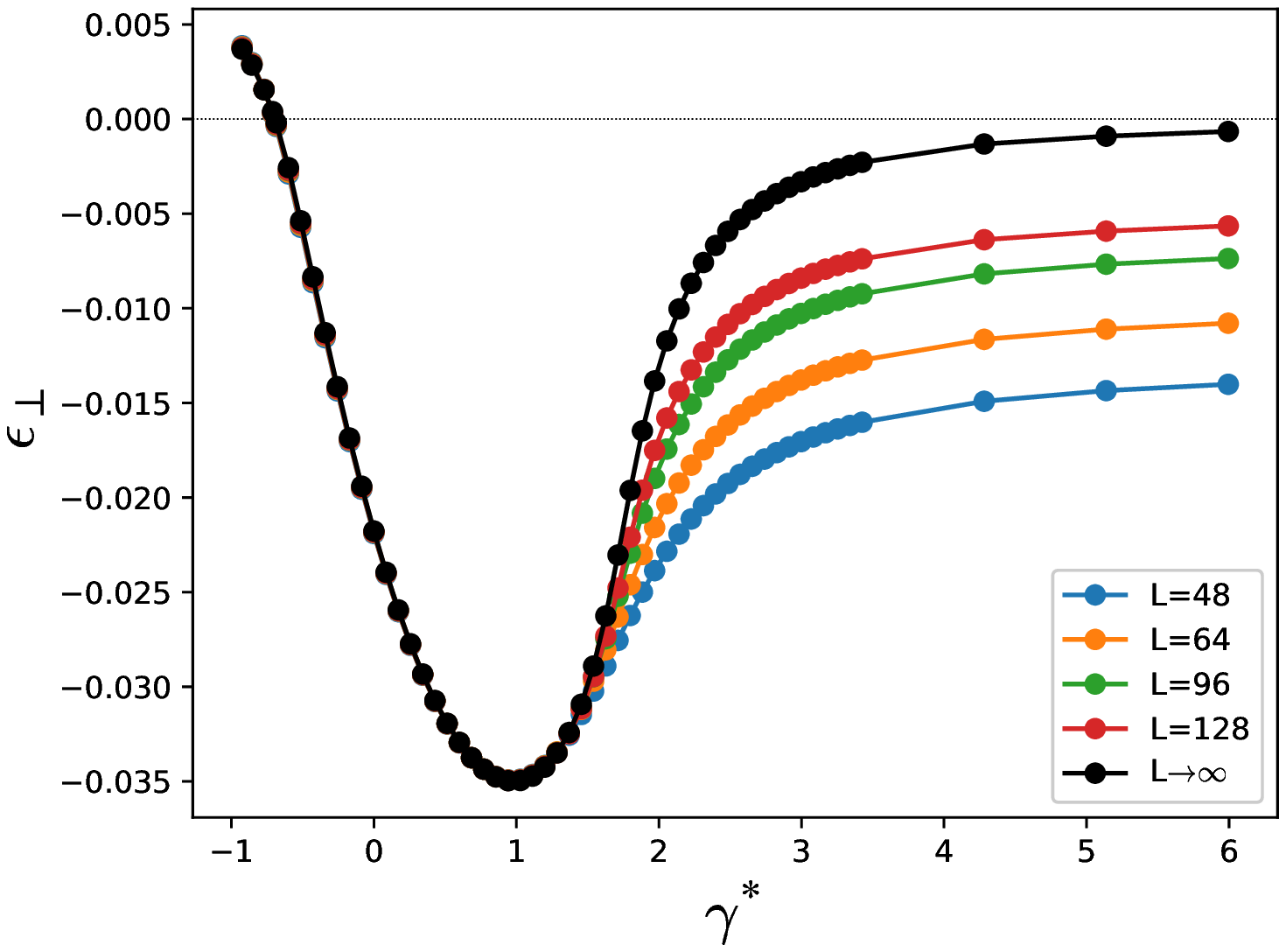}
\caption{
Various order parameters, computed as staggered averages of local ground state expectation values. Top right panel: spin chirality. Top left panel: staggered magnetization.
Bottom panels:  longitudinal and transverse spin-spin correlations. DMRG results for finite chains  and their $1/L$ extrapolations.
}
\label{fig:DMRG-order}
\end{center}
\end{figure}

\end{widetext}

In Fig.\ \ref{fig:DMRG-order} we show the various order parameters discussed in the previous Sections.
The top right panel shows the
staggered part of the chirality operator expectation value in the
ground state. These results confirm that the long-range spin chirality dimer order, already computed at
$\gamma^{\ast}=0$ in section \ref{sec:TheModel}, remains present for $\gamma^{\ast}>-1$ with a maximum at $\gamma^{\ast}\approx 1$.
The top right panel shows the $z$-component of the staggered
magnetization in the $S_{z}^{tot} = 0$ ground state. One can see
that it is strictly  zero for $-1 < \gamma^{\ast} <
\gamma^{\ast}_{c2} \sim 2.5$ and raises suddenly thereof. This
agrees with the bosonization analysis in subsection
{\ref{subsec:C2}, signaling an Ising  transition to the
antiferromagnetic ordered phase.
Finally, in the bottom  panels we show the staggered part of the
transverse  and longitudinal components of the spin exchange.
Oscillations of both components tend to zero as AFM LRO dominates
and $|\langle S^z_n\rangle|$ tends to $0.5$.

\section{Summary}

In this paper, we have studied the ground-state properties of the
one-dimensional spin $S=1/2$ XXZ Heisenberg chain with spatially
modulated Dzyaloshinskii-Moriya (DM) interaction.  Our goal was to
describe the interplay between the uniform and staggered parts of
the DM  interaction which, when acting alone, do not change the
excitation spectrum of the system. We have shown that joint effect
of the uniform and staggered components of the DM coupling opens a
possibility for formation of unconventional gapped phases in the
ground-state of the system

Depending on the effective anisotropy parameter
$\gamma^{\ast}=J_{z}/\sqrt{J^{2}+D_{0}^{2}+D_{1}^{2}}$, besides the
standard ferromagnetic at $\gamma^{\ast}\leq -1$ and gapless
Luttinger-liquid phase at $-1<\gamma^{\ast}< \gamma^{\ast}_{c1}$,
the ground state phase diagram of the model contains two
unconventional composite gapped phases. The gapped C1 phase exists
for $\gamma^{\ast}_{c1}<\gamma^{\ast} <\gamma^{\ast}_{c2}$ and is
characterized by the coexistence of LRO dimerization and alternating
spin chirality patterns, while the composite C2 phase, which is
realized at $\gamma^{\ast} > \gamma^{\ast}_{c2}>1$, is characterized
by the presence, in addition to the dimerization and alternating
spin chirality order, of long-range antiferromagnetic order.

Exploring the critical properties of the effective double
sine-Gordon theory we argue, that the transition from the LL to the
C1 phase at $\gamma^{\ast}_{c1}$ belongs to the
Berezinskii-Kosterlitz-Thouless universality class, while the
transition at $\gamma^{\ast}=\gamma^{\ast}_{c2}$ from C1 to C2 phase
is of the Ising type. Extensive DMRG results support these
statements.

\section{Acknowledgment}

We are grateful to A.A. Nersesyan and D.C. Cabra for their interest
in this work and helpful comments.

\end{document}